\documentclass[11pt,a4 paper]{article}
\usepackage{amssymb}
\usepackage{amsmath}
\usepackage[dvips]{graphicx}
\begin{document}
\thispagestyle{empty}
\noindent {\textbf{\Large  Friedmann model with viscous cosmology in modified $f(R,T)$ gravity theory}}

\vspace{1cm}

\noindent  \textbf{C. P. Singh \footnote{Corresponding author}, \;\textbf{Pankaj Kumar\;$^{2}$ }}\\

\vspace{0.5cm}

\noindent{ $^{1,2}$Department of Applied Mathematics,\\
 Delhi Technological University (Formerly Delhi College of
 Engineering)\\
 Bawana Road, Delhi-110 042, India.}\\
\texttt{ $^1$cpsphd@rediffmail.com  \\
$^2$pankaj.11dtu@gmail.com}\\

\vspace{1.5cm}
\noindent {\textbf{Abstract}} In this paper, we introduce bulk viscosity in the formalism of modified gravity theory in which the gravitational action contains a general function $f(R,T)$, where $R$ and $T$ denote the curvature scalar and the trace of the energy-momentum tensor, respectively within the framework of a flat Friedmann-Robertson-Walker model. As an equation of state for prefect fluid, we take $p=(\gamma-1)\rho$, where $0 \leq \gamma \leq 2$ and viscous term as a bulk viscosity due to isotropic model, of the form $\zeta =\zeta_{0}+\zeta_{1}H$, where $\zeta_{0}$ and $\zeta_{1}$ are constants, and $H$ is the Hubble parameter. The exact non-singular solutions to the corresponding field equations are obtained with non- viscous and viscous fluids, respectively by assuming a simplest particular model of the form of $f(R,T) = R+2f(T)$, where $f(T)=\alpha T$ ( $\alpha$ is a constant). A big-rip singularity is also observed for $\gamma<0$ at a finite value of cosmic time under certain constraints. We study all possible scenarios with the possible positive and negative ranges of $\alpha$  to analyze the expansion history of the universe. It is observed that the universe accelerates or exhibits transition from decelerated phase to accelerated phase under certain constraints of $\zeta_0$ and $\zeta_1$. We compare the viscous models with the non-viscous one through the graph plotted between scale factor and cosmic time and find that bulk viscosity plays the major role in the expansion of the universe. A similar graph is plotted for deceleration parameter with non-viscous and viscous fluids and find a transition from decelerated to accelerated phase with some form of bulk viscosity. \\

\pagebreak

\pagestyle{myheadings}
\noindent\textbf{1 Introduction}\\

 \noindent Recently, the modified theory of gravity has become one of the most popular candidates to understand the problem of dark energy. In literature, a number of modified theories have been discussed to explain early and late time expansion of the universe. In this context, $f(R)$ gravity ( $R$ being the Ricci scalar) is the simplest and most popular modification of General Relativity (GR). The $f(R)$ gravity was first introduced in [1] and later used to find a non-singular isotropic de-Sitter type cosmological solution [2](for recent reviews on $f(R)$ gravity, see refs. [3-6]). In $f(R)$ gravity, it has been suggested that the cosmic acceleration can be achieved  by replacing Einstein-Hilbert action of GR with a general function $f(R)$. Many authors [7-11] have investigated several aspects of $f(R)$ gravity in different cosmological models. The $f(R)$ gravity can produce cosmic inflation, current cosmic acceleration and behavior of dark matter. The $f(R)$ gravity is also compatible with the observational data [12-14]. Other modified gravities like scalar Gauss-Bonnet $f(G)$ gravity [15], $f(T)$ gravity [16], where $T$ is the torsion  have been proposed to overcome many problems in cosmology. Therefore, modifying the law of gravity is a possible way to explain the acceleration mechanism of the universe.\\
\indent Bertolami et al. [17] generalized $f(R)$ gravity by introducing an explicit coupling between arbitrary function of the Ricci scalar $R$ and the matter Lagrangian density $\mathcal{L}_{m}$. It provides a non-minimal coupling between matter and geometry in a more general manner at the action level. A maximal extension of the Einstein-Hilbert Lagrangian was introduced in [18], where the Lagrangian of the gravitational field was considered to be a general function of $R$ and $\mathcal{L}_{m}$.  In $f(R, \mathcal{L}_{m})$ gravity, it is assumed that all the properties of the matter are encoded in the matter Lagrangian $\mathcal{L}_{m}$.  Harko and Lobo [19] generalized this concept to the arbitrary coupling between matter and geometry. \\
\indent Recently, Harko et al. [20] have introduced another extension of GR, so called the $f(R,T)$ modified theory of gravity, where the gravitational Lagrangian is given by an arbitrary function of the Ricci scalar $R$ and the trace $T$ of the energy-momentum tensor. The authors suggested that the coupling of matter and geometry leads to a model which depends on a source term representing the variation of the energy-momentum tensor with respect to the metric. This theory has been recently introduced as modifications of Einstein's theory possessing some interesting solutions which are relevant in cosmology and astrophysics. Houndjo and co-authors [21] investigated $f(R,T)$ gravity models to reproduce the four known finite-time future  singularities. Alvarenga et al. [22] tested some $f(R,T)$ gravity models through energy conditions. Pasqua et al. [23] studied a particular model $f(R,T)=\mu R+\nu T$ which describes a quintessence-like behavior and exhibits transition from decelerated to accelerated phase. Sharif and Zubair [24] considered two forms of the energy-momentum tensor of dark components and demonstrated that the equilibrium description of thermodynamics can not be achieved at the apparent horizon of Friedmann-Robertson-Walker (FRW) universe in $f(R,T)$ gravity. Azizi [25] showed the matter threading of the wormhole may satisfy the energy conditions and the effective energy-momentum tensor is responsible for violation of the null energy condition in $f(R,T)$ gravity. Chakraborty [26] incorporated conservation of energy -momentum tensor in the field equations and discussed the energy conditions in $f(R,T)$ gravity with perfect fluid. Recently, Singh and Singh [27] have presented the reconstruction of $f(R,T)$ gravity with perfect fluid in the framework of FRW model for two well-known scale factors.\\
\indent In recent years the observations like type Ia supernovae [28-31], cosmic microwave background [32, 33] and large scale structure [34, 35] describe about an accelerated expansion of the universe. In most of the cosmological models, the content of the universe has been considered as a perfect fluid. It is important to investigate more realistic models that take into account dissipative processes due to viscosity. In a homogeneous and isotropic universe bulk viscosity is the unique viscous effect capable to modify the background dynamics. It is known that when neutrino decoupling occurred, the matter behaved like a viscous fluid in the early stage of the universe. There were remarkable cosmological applications of viscous imperfect fluids already in the 1970s [36]. In the context of inflation, it has been known since long time ago that an imperfect fluid with bulk viscosity can produce an acceleration without the need of a cosmological constant or some scalar field.  An inflationary epoch driven by bulk viscous pressure has also been proposed in the 1980s [37]. All these works have analyzed the role played by bulk viscosity in the early universe. \\
\indent In the framework of homogeneous and isotropic universe, for a sufficiently large bulk viscosity, the effective pressure becomes negative and hence it can explain late time acceleration of the universe. The dark energy phenomena as an effect of the bulk viscosity in the cosmic media has been first investigated in refs. [38]. All these cited works are considered in pioneer papers on cosmological bulk viscosity, but it is also worth noting some recent applications of viscous fluids as candidates for dark matter [39], dark energy [40] or unified models [41], i.e., when a single substance plays the role of both dark matter and dark energy simultaneously. Indeed, it has been shown that for an appropriate viscosity coefficient, an accelerating cosmology can be achieved without the need of a cosmological constant [42, 43]. In refs.[44-48], the bulk viscous cosmological models have been studied in various aspects. In an accelerated expanding universe it may be natural to assume the possibility that the expansion process is actually a collection of states out of thermal equilibrium in a small fraction of time due to existence of a bulk viscosity. Hence, FRW cosmology may be modelled as a bulk viscosity within a thermodynamical approach. A well known result of the FRW cosmological solutions, corresponding to the universe filled with perfect fluid and bulk viscous stresses, is the possibility of violating dominant energy condition(DEC). At the late times, since we do not know the nature of the universe content (dark matter and dark energy components) very clearly, concerning about the bulk viscosity is reasonable and practical. To our knowledge, such possibility has been investigated only in the context of the primordial universe, concerning also the search of non-singular models. But many investigations show that the viscous pressure can play the role of an agent that drives the present acceleration of the universe. The motivation of the present work is to drive the present acceleration using the bulk viscous pressure within the cosmic fluid instead of any dark energy component in modified $f(R,T)$ gravity theory.\\
\indent In the present paper, we study FRW model with bulk viscosity in modified $f(R,T)$ gravity theory and investigate the effects of bulk viscosity in explaining the early and late time acceleration of the universe. The model contains the perfect fluid with bulk viscosity of the form $\zeta =\zeta_{0}+\zeta_{1}H$, where $\zeta_{0}$ and $\zeta_{1}$ are constants and $H$ is the Hubble parameter. The exact solutions of field equations are obtained with constant and varying bulk viscosity by assuming a simplest particular form of $f(R,T) = R+2f(T)$, where $f(T)=\alpha T$. We study all possible scenarios according to the values of $\alpha$ and under the constraints of $\zeta_{0}$ and $\zeta_1$ to analyze the behavior of the scale factor, matter density and discuss the expansion history of the universe. We find cosmological solutions which exhibit a big rip singularity under certain constraints. Therefore, the negative pressure generated by the bulk viscosity can not avoid the dark energy of the universe to be phantom. \\
\indent The paper is organized as follows. In section 2 we present the brief review of the modified $f(R,T)$ gravity theory as proposed by Harko et al. [20]. Section 3 presents the cosmological model and its field equations with a bulk viscous fluid. Section 4 is divided into two subsections 4.1 and 4.2. In 4.1 we present the solution with constant bulk viscosity for two cases 4.1.1 and 4.1.2 and the solutions with time-dependent bulk viscosity is presented in 4.2.1, 4.2.2 and 4.2.3 of section 4.2. In Section 5 we summarize our results in detail. \\

\noindent \textbf{2 Brief review of modified $f(R,T)$ gravity theory}\\

\noindent The $f(R,T)$ theory is a modified theory of gravity, in which the Einstein-Hilbert Lagrangian, i.e., $R$ is replaced by an arbitrary function of the scalar curvature $R$ and the trace $T$ of energy-momentum tensor.  In [20], the following modification of Einstein's theory is proposed in the unit $8\pi G=1=c$.
\begin{equation}
S=\frac{1}{2}\int d^{4}x\sqrt{-g}[f(R,T)+2\mathcal{L}_{m}],
\end{equation}
\noindent where $g$ is the determinant of the metric tensor $g_{\mu\nu}$ and $\mathcal{L}_{m}$ is the matter Lagrangian density.  The energy-momentum tensor $T_{\mu\nu}$, defined from matter Lagrangian density $\mathcal{L}_{m}$ is given by
 \begin{equation}
  T_{\mu\nu}=-\frac{2}{\sqrt{-g}}\frac{\delta(\sqrt{-g}\;\mathcal{L}_m)}{\delta g^{\mu\nu}},
\end{equation}
\noindent  and its trace by $T=g^{\mu\nu}T_{\mu\nu}$.  Assuming  the matter Lagrangian density $\mathcal{L}_m$ depends only on the metric tensor components $g_{\mu\nu}$, not on its derivatives, we obtain
\begin{equation}
  T_{\mu\nu}=g_{\mu\nu}\mathcal{L}_m-2\frac{\partial \mathcal{L}_m}{\partial g^{\mu\nu}}.
\end{equation}
\noindent The equations of motion by varying the action (1) with respect to metric tensor are given by [20]
\begin{equation}
f_{R}(R,T)R_{\mu\nu}-\frac{1}{2}f(R,T)g_{\mu\nu}+(g_{\mu\nu}\square-\nabla_\mu\nabla_\nu)f_{R}(R,T)=
T_{\mu\nu}-f_{T}(R,T)T_{\mu\nu}-f_{T}(R,T)\circleddash_{\mu\nu},
\end{equation}
\noindent where $f_R$ and $f_T$ denote the derivatives of $f(R,T)$ with respect to $R$ and $T$, respectively. Here, $\nabla_\mu$ is covariant derivative and $\square\equiv\nabla_\mu\nabla^\mu$ is the d'Alembert operator and $\circleddash_{\mu\nu}$ is defined by
\begin{equation}
  \circleddash_{\mu\nu}\equiv g^{\alpha\beta}\frac{\delta T_{\alpha\beta}}{\delta g^{\mu\nu}}.
\end{equation}
\noindent Using (3) into (5), we obtain
\begin{equation}
\circleddash_{\mu\nu}=-2T_{\mu\nu}+g_{\mu\nu}\mathcal{L}_m-2g^{\alpha\beta}\frac{\partial^2\mathcal{L}_m}{\partial g^{\mu\nu}\partial g^{\alpha\beta}}.
\end{equation}
\noindent The equations of $f(R,T)$ gravity are much more complicated with respect to the ones of General Relativity even for FRW metric. For this reason many possible form of $f(R,T)$, for example, $f(R,T)= R+2f(T)$,  $f(R,T)$= $\mu f_1(R)+\nu f_2(T)$, where $f_1(R)$ and $f_2(T)$ are arbitrary functions of $R$ and $T$, and $\mu$ and $\nu$ are real constants, respectively [20-23],  and $f(R,T)=R\;f(T)$ [26], etc., have been proposed to solve the modified field equations. In this paper we consider the following simplest particular model (as one considered in [20]):
\begin{equation}
f(R,T)=R+2f(T),
\end{equation}
\noindent i.e. the action is given by the same Einstein-Hilbert one plus a function of $T$. The term $2f(T)$ in the gravitational action modifies the gravitational interaction between matter and curvature. Using (7), one can re-write the gravitational field equations defined in (4) as
\begin{equation}
R_{\mu\nu}-\frac{1}{2}R g_{\mu\nu}=T_{\mu\nu}-2 (T_{\mu\nu}+\circleddash_{\mu\nu})f'(T)+f(T) g_{\mu\nu},
\end{equation}
\noindent which is considered as the field equations of $f(R,T)$ gravity. Here, a prime stands for derivative of $f(T)$ with respect to $T$. The assumption (7) is particularly interesting choice since, for $p=0$ one has $T=\rho$ and, by choosing $f(T)=\alpha T$ where $\alpha$ is a constant, one can construct a model with an effective cosmological constant. In order to compare (8) with Einstein's, we find that the gravitational field equations (8) can be recast in such a form that the higher order corrections coming both from the geometry , and from matter -geometry coupling , provide an energy-momentum tensor of geometrical and matter origin, describing an effective source term on right hand side of (8).\\
\indent The main issue now arises on the content of the universe through the energy-momentum tensor and consequently on the matter Lagrangian $\mathcal{L}_m$ and the trace of the energy-momentum tensor.\\

\noindent \textbf{3 Metric and field equations}\\

\noindent We assume a spatially homogeneous and isotropic flat Friedmann-Robertson-Walker (FRW) metric
\begin{equation}
ds^{2}=dt^{2}-a^{2}(t)(dx^{2}+dy^{2}+dz^{2}),
\end{equation}
\noindent where $a(t)$ is the cosmic scale factor.\\
\indent In comoving coordinates, the components of the four-velocity $u^{\mu}$ are $u^{0}=1$, $u^{i}=0$. With the help of the projection tensor $h_{\mu\nu}=g_{\mu\nu}+u_{\mu}u_{\nu}$, we have the energy momentum tensor for a viscous fluid [49, 50]
\begin{equation}
T_{\mu\nu}=\rho u_{\mu}u_{\nu}-\bar{p}h_{\mu\nu},
\end{equation}
\noindent where $\bar{p}$ denotes the effective pressure. In the first order thermodynamics theory of Eckart [51], $\bar{p}$ is given by
 \begin{equation}
 \bar{p}=p-3\zeta H.
 \end{equation}
\noindent Thus, for large $\zeta$ it is possible for negative pressure term to dominate and an accelerating cosmology to ensue. Here, $H=\dot{a}/a$ is Hubble parameter, where an overdot means differentiation with respect to $t$, and $\rho$, $p$ and $\zeta$ are the energy density, the isotropic pressure and coefficient of bulk viscosity, respectively. Therefore, the Lagrangian density may be chosen as $\mathcal{L}_m=-\bar{p}$, and  the tensor $\circleddash_{\mu\nu}$ in (6) is given by
\begin{equation}
\circleddash_{\mu\nu}=-2T_{\mu\nu}-\bar{p}g_{\mu\nu}.
\end{equation}
Using (10) and (12), the field equations (8) for bulk viscous fluid become
\begin{equation}
R_{\mu\nu}-\frac{1}{2}R g_{\mu\nu}=T_{\mu\nu}+2f^{\prime}(T)T_{\mu\nu}+\left(2\bar{p}f^{\prime}(T) +f(T)\right)\; g_{\mu\nu}.
\end{equation}
\noindent The field equations (13) with the particular choice of the function $f(T)=\alpha T$, where $\alpha$ is a constant [see, Harko et al.[20]] for the metric (9) yield
\begin{equation}
3H^{2}=\rho + 2\alpha(\rho+\bar{p})+\alpha T,
\end{equation}
\begin{equation}
2\dot{H}+3H^{2}=-\bar{p}+\alpha T,
\end{equation}
\noindent where $T= \rho-3\bar{p}$. We have two independent equations (14) and (15), and four unknown variables, namely $H$, $\rho$, $p$ and $\zeta$ to be solved as functions of time. In the following section we choose equation of state and bulk viscosity and try to solve for $H$.\\

\noindent \textbf{4 Solution of field equations}\\

\noindent From (14) and (15) we get a single evolution equation for $H$:
\begin{equation}
2\dot{H}+(1+2\alpha)(\rho+p)-3(1+2\alpha)\zeta H=0.
\end{equation}
Thus, if an equation of state (EoS) connecting $p$ and $\rho$ is chosen in the form
\begin{equation}
p=(\gamma-1)\rho,
\end{equation}
\noindent where $\gamma$ is a constant known as the EoS parameter lying in the range $0\leq \gamma \leq 2$, then equation (16) can be solved for any particular choice of $\zeta$.  \\
\indent Let us assume the general bulk viscosity $\zeta$ of the form [48,52,53]:\\
\begin{equation}
\zeta=\zeta_{0}+\zeta_{1}H,
\end{equation}
where $\zeta_{0}$ and $\zeta_{1}$ are two constants conventionally. The motivation of considering this bulk viscosity is that by fluid mechanics. We know that the transport / viscosity phenomenon is involved with the ``velocity" $\dot{a}$, which is related to the scalar expansion $\theta=3\dot{a}/a$. Both $\zeta=\zeta_{0}$ (constant) and $\zeta \propto \theta$ are separately considered by many authors. Therefore, a linear combination of the two are more general. \\
\noindent Using (11), (17) and (18) into (14), we have
\begin{equation}
\rho=\frac{3H[(1-\alpha \zeta_{1})H-\alpha \zeta_{0}]}{1+4\alpha-\alpha\gamma}.
\end{equation}

\noindent \textbf{4.1 Cosmology with non-viscous fluid}\\

\noindent In this case, where $\zeta=0$, equation (16), with the help of (17), (18) and (19), reduces to
\begin{equation}
\dot{H}+\frac{3}{2}\frac{\gamma(1+2\alpha)H^2}{(1+4\alpha-\alpha\gamma)}=0.
\end{equation}
On solving (20) for $\gamma\neq 0$, we get
\begin{equation}
H=\frac{1}{[C+\frac{3(1+2\alpha)\gamma}{2(1+4\alpha-\alpha\gamma)}\;t]},
\end{equation}
where $C$ is a constant of integration. Using $H = \dot {a}/{a}$, equation (21) gives the power-law expansion for the scale factor of the form
\begin{equation}
a=D\left[C+ \frac{3\gamma(1+2\alpha)}{2(1+4\alpha-\alpha\gamma)}\;t \right]^{\frac{2(1+4\alpha-\alpha\gamma)}{3\gamma(1+2\alpha)}},\;\;\;\;( \alpha\neq -1/2),
\end{equation}
where $D$ is another constant of integration. This scale factor can be rewritten as
\begin{equation}
a=a_{0}\left[1+ \frac{3\gamma(1+2\alpha)H_{0}}{2(1+4\alpha-\alpha\gamma)}\;t \right]^{\frac{2(1+4\alpha-\alpha\gamma)}{3\gamma(1+2\alpha)}},
\end{equation}
where $H=H_{0}>0$ at $t=t_{0}$. The cosmic time $t_{0}$ corresponds to the time where dark component begins to become dominant. The energy density is given  by
\begin{equation}
\rho=\rho_{0}\left[1+\frac{3\gamma(1+2\alpha)H_{0}}{2(1+4\alpha-\alpha\gamma)}\;t \right]^{-2},
\end{equation}
where $\rho_{0}=3H^{2}_{0}/(1+4\alpha-\alpha \gamma)$. For $\gamma<0 $ we get a big rip singularity at finite time $t_{br} = -2(1+4\alpha-\alpha\gamma)/3\gamma(1+2\alpha)H_{0}>t_{0}$ as the scale factor and energy density tend to infinite at this time.\\
\indent The cosmological inflation or the accelerated expansion of the universe is characterized by the deceleration parameter $q$ defined by $q=-\frac{\ddot{a}a}{\dot{a}^{2}}$. In this case, we get
\begin{equation}
q=\frac{3\gamma(1+2\alpha)}{2(1+4\alpha-\alpha\gamma)}-1,
\end{equation}
which is constant through out the evolution of the universe. As we know that $q>0$ determines expansion of universe with decelerated rate, $q<0$ describes the accelerated expansion of the universe and $q=0$ gives the coasting or marginal inflation. Thus, for suitable values of $\alpha$ and $\gamma$ we can obtain decelerated and accelerated expansion of the universe. In this case, the model does not exhibit phase transition due to constant value of $q$.\\
\indent For $\gamma = 0$ i.e. $p=-\rho$, equation (20) gives $H=H_{0}$, which corresponds to $a = a_{0}e^{H_{0}t}$ i.e. de-Sitter type expansion of the universe. Both $\rho$ and $p$ are constant and $q=-1$ through out the evolution of the universe.\\

\noindent \textbf{4.2 Cosmology with viscous fluid}\\

\noindent On the thermodynamical grounds, $\zeta$ in (18) is conventionally chosen to be a positive quantity and may depend on the cosmic time $t$, or the scale factor $a$, or the energy density $\rho$. Therefore, different forms of viscosity can be used to make (16) solvable numerically or exactly. We investigate  in the following some different choices for $\zeta$.\\

\noindent\textbf{4.2.1 Solution with constant bulk viscosity}\\

\noindent From bulk viscosity point of view, the simplest case is thought to be a constant bulk viscosity. Therefore, assuming $\zeta_{1}=0$ in (18), we get $\zeta=\zeta_{0}$.  In this case (19) reduces to
\begin{equation}
\rho=\frac{3H^{2}-3\alpha\zeta_{0}H}{(1+4\alpha-\alpha\gamma)}.
\end{equation}
\noindent Substituting (17) and (26) into (16), we get
\begin{equation}
\dot{H}+\frac{3}{2}\frac{\gamma(1+2\alpha)H}{(1+4\alpha-\alpha\gamma)}\left[H-\frac{\zeta_{0}(1+4\alpha)}
{\gamma}\right]=0.
\end{equation}
\noindent In what follows we solve (27) for $\gamma \neq 0$ and $\gamma=0$ separately.\\

\noindent {\it {Case I}: Solution for $\gamma \neq 0$}\\

\noindent Solving (27) for $\gamma \neq 0$, we find
\begin{equation}
H=\frac{e^{\frac{3}{2}\frac{(1+2\alpha)(1+4\alpha)\zeta_{0}}{(1+4\alpha-\alpha\gamma)}\;t}}{c_{0}+\frac{\gamma}
{(1+4\alpha)\zeta_{0}}\;e^{\frac{3}{2}\frac{(1+2\alpha)(1+4\alpha)\zeta_{0}}{(1+4\alpha-\alpha\gamma)}\;t}},
\end{equation}
\noindent where $c_{0}$ is a constant of integration. Using $H=\dot{a}/a$, the scale factor in terms of $t$ is given by
\begin{equation}
a=c_{1}\left[c_{0}+\frac{\gamma}
{(1+4\alpha)\zeta_{0}}\;e^{\frac{3}{2}\frac{(1+2\alpha)(1+4\alpha)\zeta_{0}}{(1+4\alpha-\alpha\gamma)}\;t}
\right]^{\frac{2}{3}\frac{(1+4\alpha-\alpha\gamma)}{\gamma(1+2\alpha)}},
\end{equation}
\noindent where $c_{1} > 0$ is another integration constant. This scale factor may be rewritten as \\
\begin{equation}
a(t)=a_{0}\left[1+\frac{\gamma H_{0}}
{(1+4\alpha)\zeta_{0}}\left(e^{\frac{3}{2}\frac{(1+2\alpha)(1+4\alpha)\zeta_{0}}{(1+4\alpha-\alpha\gamma)}\;t}-1\right)
\right]^{\frac{2}{3}\frac{(1+4\alpha-\alpha\gamma)}{\gamma(1+2\alpha)}}.
\end{equation}
\indent The energy density can be calculated as
\begin{eqnarray}
\rho &=&\frac{3H_{0}}{(1+4\alpha-\alpha \gamma)}\left[\frac{e^{\frac{3}{2}\frac{(1+2\alpha)(1+4\alpha)\zeta_{0}}{(1+4\alpha-\alpha\gamma)}\;t}}{1+\frac{\gamma H_{0}}
{(1+4\alpha)\zeta_{0}}\;\left(e^{\frac{3}{2}\frac{(1+2\alpha)(1+4\alpha)\zeta_{0}}{(1+4\alpha-\alpha\gamma)}\;t}-1\right)}\right]\nonumber\\
& &\left[\frac{H_{0}\;e^{\frac{3}{2}\frac{(1+2\alpha)(1+4\alpha)\zeta_{0}}
{(1+4\alpha-\alpha\gamma)}\;t}}{1+\frac{\gamma H_{0}}
{(1+4\alpha)\zeta_{0}}\;\left(e^{\frac{3}{2}\frac{(1+2\alpha)(1+4\alpha)\zeta_{0}}{(1+4\alpha-\alpha\gamma)}\;t}-1\right)}-\alpha\; \zeta_{0}\right].
\end{eqnarray}
\noindent For $0 \leq\gamma\leq 2$, viscous solution satisfies the dominant energy condition (DEC), i.e., $\rho+p\geq0$. If $\gamma < 0$ we have a big rip singularity at a finite value of cosmic time
\begin{equation}
t_{br}=\frac{2(1+4\alpha-\alpha \gamma )}{3(1+2\alpha)(1+4\alpha)\zeta_{0}}\;ln \left(1-\frac{(1+4\alpha)\zeta_{0}}{\gamma H_{0}}\right) > t_{0}.
\end{equation}
\noindent One may observe that there is a violation of DEC. The energy density grows up to infinity at a finite time $t>t_{0}$, which leads to a big rip singularity characterized by the scale factor and Hubble parameter blowing up to infinity at this finite time. Therefore, there are cosmological models with viscous fluid which present in the development of this sudden future singularity. \\
\indent The deceleration parameter is given by
\begin{equation}
q=\frac{\frac{3}{2}\frac{(1+2\alpha)}{(1+4\alpha-\alpha \gamma)}[\gamma-\frac{(1+4\alpha)\zeta_{0}}{H_{0}}]}
{e^{\frac{3}{2}\frac{(1+2\alpha)(1+4\alpha)\zeta_{0}}{(1+4\alpha-\alpha\gamma)}\;t}}-1,
\end{equation}
\noindent which is time-dependent in contrast to perfect fluid. Thus,  the constant bulk viscous coefficient generates time-dependent $q$ which may also describes the transition phases of the universe along with deceleration or acceleration of the universe. Let us observe the variation of $q$ with bulk viscous coefficient $\zeta_{0}$ for various ranges of $\alpha$ in different phases of evolution of the universe for $\gamma >0$, which are  presented in the following tables 1-3.\\
\begin{center}
\textbf{Table. 1} : \textsf{Variation of $q$ for $\gamma=2/3$.}\\
\vspace{0.2cm}
\footnotesize{\begin{tabular}{|c|c|c|c|}
  \hline
  Range of $\alpha $& Constraints on $\zeta_{0}$ & $q$  &Evolution of Universe\\ \hline
  $\alpha > 0$ & for all $\zeta_{0}>0$ & negative &  accelerated expansion  \\ \hline
  $-0.25<\alpha<0$  & $0<\zeta_{0}<\frac{-8\alpha H_{0}}{9(1+2\alpha)(1+4\alpha)}$ & +ve to -ve &  transition from dec. to acc.   \\
  & $\zeta_{0}\geq\frac{-8\alpha H_{0}}{9(1+2\alpha)(1+4\alpha)}$ &negative &  accelerated expansion  \\ \hline
  $-0.30<\alpha\leq-0.25$ & for all $\zeta_{0}>0$ & positive &  decelerated expansion  \\ \hline
$-0.50\leq \alpha<-0.30$ & for all $\zeta_{0}>0$ & negative &  accelerated  expansion  \\ \hline
 $\alpha<-0.50$  & $0<\zeta_{0}<\frac{-8\alpha H_{0}}{9(1+2\alpha)(1+4\alpha)}$ & -ve to +ve &  transition from acc. to dec. \\
  &$\zeta_{0}\geq\frac{-8\alpha H_{0}}{9(1+2\alpha)(1+4\alpha)}$ & positive
&  decelerated expansion  \\ \hline
\end{tabular}}
\end{center}
\vspace{0.6cm}
\begin{center}
\textbf{Table. 2} : \textsf{Variation of $q$ for $\gamma=4/3$.}\\
\vspace{0.2cm}
\footnotesize{\begin{tabular}{|c|c|c|c|}
  \hline
  Range of $\alpha $ & Constraints on $\zeta_{0}$ & $q$ &Evolution of Universe \\ \hline
  $\alpha > -0.25$ ($\alpha \neq 0$) & $0<\zeta_{0}<\frac{2(3+4\alpha)H_{0}}{9(1+4\alpha)(1+2\alpha)} $ & +ve to -ve & transition from dec. to acc. \\
  & $\zeta_{0}\geq \frac{2(3+4\alpha)H_{0}}{9(1+4\alpha)(1+2\alpha)} $ & negative & accelerated expantion\\ \hline
 $-0.375<\alpha\leq-0.25$ & for all $\zeta_{0}>0$ & positive & decelerated expansion  \\ \hline
$-.50\leq \alpha<-0.375$ & for all $\zeta_{0}>0$ & negative & accelerated  expansion  \\ \hline
$\alpha<-0.50$  & $0<\zeta_{0}<\frac{2(3+4\alpha)H_{0}}{9(1+4\alpha)(1+2\alpha)} $ & -ve to +ve & transition from acc. to dec.\\
  &$\zeta_{0}\geq\frac{2(3+4\alpha)H_{0}}{9(1+4\alpha)(1+2\alpha)}$ & positive
& decelerated expansion\\ \hline
\end{tabular}}
\end{center}
\vspace{0.2cm}
\begin{center}
\textbf{Table. 3} : \textsf{Variation of $q$ for $\gamma=1$.}\\
\vspace{0.2cm}
\footnotesize{\begin{tabular}{|c|c|c|c|}
  \hline
  Range of $\alpha $ & Constraints on $\zeta_{0}$ & $q$ &Evolution of Universe \\ \hline
  $\alpha > -0.25$ ($\alpha \neq 0$) & $0<\zeta_{0}<\frac{H_{0}}{3(1+2\alpha)(1+4\alpha)} $ & +ve to -ve & transition from dec. to acc. \\
  & $\zeta_{0}\geq \frac{H_{0}}{3(1+2\alpha)(1+4\alpha)} $ & negative & accelerated expantion\\ \hline
 $-0.33<\alpha\leq-0.25$ & for all $\zeta_{0}>0$ & positive & decelerated expansion  \\ \hline
$-0.50\leq \alpha<-0.33$ & for all $\zeta_{0}>0$ & negative & accelerated  expansion  \\ \hline
$\alpha<-0.50$  & $0<\zeta_{0}<\frac{H_{0}}{3(1+2\alpha)(1+4\alpha)} $ & -ve to +ve & transition from acc. to dec.\\
  &$\zeta_{0}\geq\frac{H_{0}}{3(1+2\alpha)(1+4\alpha)}$ & positive
& decelerated expansion\\ \hline
\end{tabular}}
\end{center}

\vspace{0.1cm}

\noindent We observe from the above tables 1-3 that the universe accelerates through out the evolution when $\alpha > 0$ for any $\zeta_{0}> 0 $ in inflationary phase where as it shows transition from decelerated phase to accelerated phase when $\alpha > -0.25$ for smaller values of $\zeta_{0}$ and acceleration for larger values of $\zeta_{0}$ in radiation and matter-dominated phases. It is to be noted here that the larger values of $\zeta_{0}$ makes the effective pressure more negative to accelerate the universe through out the evolution. We find that the universe decelerates in $-0.30<\alpha\leq-0.25$ for $\gamma=2/3$, in $-0.375<\alpha\leq-0.25$ for $\gamma=4/3$ and in $-0.33<\alpha\leq-0.25$ for $\gamma=1$ for all $\zeta_{0}>0$. Further, we find that the universe accelerates in $-0.50\leq \alpha<-0.30$ for $\gamma=2/3$, in $-0.50\leq \alpha<-0.375$ for $\gamma=4/3$ and in $-0.50\leq \alpha<-0.33$ for $\gamma=1$ for all $\zeta_{0}>0$ . When $\alpha <-0.50$, then model shows transition from acceleration to deceleration phase for small values of $\zeta_{0}$ and decelerates for large values of $\zeta_{0}$. In conclusion we can say that the universe accelerates or shows transition from decelerated phase to accelerated phase for $\alpha > -0.25$ with constant viscous term in all phases of its evolution.\\
\indent For $\gamma < 0$, where the solution has a big rip singularity, $q$ is always negative for $\alpha>0$ during any cosmic time.\\

\noindent{\it{Case II}: Solution for $\gamma =0$}\\

\noindent In this case, (27) reduces to
\begin{equation}
\dot{H}-\frac{3}{2}(1+2\alpha)\zeta_{0}H=0,
\end{equation}
\noindent which gives the solution for $H$ in terms of $t$ as
\begin{equation}
H= c_{2} e^{\frac{3}{2}(1+2\alpha)\zeta_{0}t},
\end{equation}
\noindent where $c_{2}>0$ is a constant of integration. The scale factor in terms of t is given by\\
\begin{equation}
a= c_{3} e^{\frac{2c_{2}}{3(1+2\alpha)\zeta_{0}}e^{\frac{3}{2}(1+2\alpha)\zeta_{0}t}},
\end{equation}
\noindent where $c_{3}>0$ is another constant of integration. This scale factor may be rewritten as \\
\begin{equation}
a= a_{0} e^{\frac{2H_{0}}{3(1+2\alpha)\zeta_{0}}(e^{\frac{3}{2}(1+2\alpha)\zeta_{0}t}-1)}.
\end{equation}
\noindent We find that the scale factor shows the superinflation in the presence of constant bulk viscous coefficient where as it has de-Sitter expansion in non-viscous case. The energy density is given by
\begin{equation}
\rho=\frac{3H_{0}e^{\frac{3}{2}(1+2\alpha)\zeta_{0}t}}{(1+4\alpha)}\left[H_{0}e^{\frac{3}{2}(1+2\alpha)\zeta_{0}t}-\alpha \zeta_{0}\right].
\end{equation}
\noindent We observe that $\rho$ varies with time in contrast to perfect fluid solution where it is constant. Both $a(t)$ and  $\rho$ tend to constant at $t=0$ and tend to infinity at $t \rightarrow\infty$. In this case the deceleration parameter is given by
\begin{equation}
q = -\frac{3\zeta_{0}(1+2\alpha)}{2H_{0}}\;e^{-\frac{3}{2}(1+2\alpha)\zeta_{0}t}-1,
\end{equation}\\
\noindent which is  time-dependent. We study the variation of $q$ with bulk viscous coefficient $\zeta_{0}$ for various ranges of $\alpha$, which are summarized in table 4. From table 4 we observe that $q$ is always negative for $\alpha \geq -0.50$ and the universe accelerates for all values of $\zeta_{0}>0$. For $\alpha<-0.50$, the value of $q$ varies from negative to positive, that is, the universe shows transition from accelerated phase to decelerated phase for small values of $\zeta_{0}$ and always decelerates for large values of $\zeta_{0}$. \\
\begin{center}
\textbf{Table. 4} : \textsf{Variation of $q$ for $\gamma=0$.}\\
\vspace{0.2cm}
\footnotesize{\begin{tabular}{|c|c|c|c|}
  \hline
  Range of $\alpha $ & Constraints on $\zeta_{0}$ & $q$ &Evolution of Universe \\ \hline
  $\alpha \geq -0.50$ ($\alpha \neq 0$) & for all $\zeta_{0}>0$  & negative & accelerated  expansion  \\ \hline
$\alpha<-0.50$  & $0<\zeta_{0}<\frac{-2H_{0}}{3(1+2\alpha)} $ & -ve to +ve & transition from acc. to dec.\\
  &$\zeta_{0}\geq\frac{2H_{0}}{3(1+2\alpha)}$ & positive & decelerated expansion\\ \hline
\end{tabular}}
\end{center}

\vspace{0.4cm}

\noindent \textbf{4.2.2 Solution with variable bulk viscosity}\\

\noindent In this section, we present the solution for $\zeta_{0}=0$ and  $\zeta_{0}\neq 0$.\\

\noindent {\it{Case I}:  $\zeta_{0}=0$}\\

\noindent For $\zeta_{0}=0$, the form of bulk viscosity assumed in (18) reduces to $\zeta=\zeta_{1} H$ [54]. This is the most interesting case. As $\zeta$ is assumed to be positive, for this the constant $\zeta_1$ must be positive.\\
\indent  Using (17) and (19) into (16) we get\\
\begin{equation}
\dot{H}+\frac{3}{2}\frac{(1+2\alpha)(\gamma-\zeta_{1}-4\alpha\zeta_{1})}{(1+4\alpha-\alpha\gamma)}
H^2=0.
\end{equation}\\
Solving (40), the Hubble parameter can be obtained in terms of $t$ for any value of $\gamma$ in the range $0\leq \gamma \leq 2$, except $\gamma\neq (1+4\alpha) \zeta_{1}$ and $\alpha\neq-1/2 $ as
\begin{equation}
H=\frac{1}{[\frac{3(1+2\alpha)(\gamma-\zeta_{1}-4\alpha\zeta_{1})}{2(1+4\alpha-\alpha\gamma)}\;t+c_{4}]},
\end{equation}
where $c_{4}$ is a constant of integration. Correspondingly, the scale factor in terms of t is given by\\
\begin{equation}
a=c_{5}\left[c_{4}+ \frac{3(1+2\alpha)(\gamma-\zeta_{1}-4\alpha\zeta_{1})}{2(1+4\alpha-\alpha\gamma)}\;t \right]^{\frac{2(1+4\alpha-\alpha\gamma)}{3(1+2\alpha)(\gamma-\zeta_{1}-4\alpha\zeta_{1})}},\;\;
\end{equation}
where $c_{5}>0$ is another constant of integration. We find that the form $\zeta=\zeta_1H$ yields a power-law expansion for the scale factor. This scale factor may be rewritten as \\
\begin{equation}
a=a_{0}\left[1+ \frac{3(1+2\alpha)(\gamma-\zeta_{1}-4\alpha\zeta_{1})H_{0}}{2(1+4\alpha-\alpha\gamma)}\;(t-t_0) \right]^{\frac{2(1+4\alpha-\alpha\gamma)}{3(1+2\alpha)(\gamma-\zeta_{1}-4\alpha\zeta_{1})}},
\end{equation}
where $H=H_0$ at $t=t_0$ and $t_0$ corresponds to the time where dark component begins to dominant, i.e., describes the present value. The energy density and bulk viscous coefficient are respectively given by
\begin{equation}
\rho=\frac{3(1-\alpha\zeta_1)H^{2}_{0}}{(1+4\alpha-\alpha \gamma)}\left[1+\frac{3(1+2\alpha)(\gamma-\zeta_{1}-4\alpha\zeta_{1})H_{0}}{2(1+4\alpha-\alpha\gamma)}\;(t-t_0) \right]^{-2},
\end{equation}
\begin{equation}
\zeta=\zeta_{1}H_{0}\left[1+\frac{3(1+2\alpha)(\gamma-\zeta_{1}-4\alpha\zeta_{1})H_{0}}{2(1+4\alpha-\alpha\gamma)}\;(t-t_0) \right]^{-1}.
\end{equation}
In this case we get deceleration parameter as\\
\begin{equation}
q=\frac{3(1+2\alpha)(\gamma-\zeta_{1}-4\alpha\zeta_{1})}{2(1+4\alpha-\alpha\gamma)}-1,
\end{equation}
\noindent which is constant. The sign of $q$ depends on the values of parameters $\gamma$, $\alpha$ and $\zeta_{1}$. It is always negative for $\gamma\leq0$, $\alpha>0$ and $\zeta_{1}>0$. \\
\indent If we demand to have the occurrence of a big rip singularity in the future then we have the following constraints on the parameters $\alpha$, $\zeta_1$ and $\gamma$
\begin{equation}
\zeta_{1}(1+4\alpha)> \gamma,
\end{equation}
which leads the scale factor and energy density tending to infinity at a finite time
\begin{equation}
t_{br}=\frac{2(1+4\alpha-\alpha\gamma)}{3(1+2\alpha)(\zeta_1+4\alpha\zeta_1-\gamma)}H^{-1}_{0}.
\end{equation}
We can observe from (43) and (44) that energy density of the dark component increases with scale factor for $\zeta_{1}(1+4\alpha)> \gamma$.\\

\noindent {{\it{Case II}:  $\zeta_{0}\neq 0$}\\

\noindent Using (17), (18) and (19) into (16) we get
\begin{equation}
\dot{H}+\frac{3}{2}\frac{(1+2\alpha)(\gamma-\zeta_{1}-4\alpha\zeta_{1})}{1+4\alpha-\alpha\gamma}H
\left[H-\frac{(1+4\alpha)\zeta_{0}}{\gamma-\zeta_{1}-4\alpha\zeta_{1}}\right]=0.
\end{equation}\\
\noindent Solving (49), we find the following solution for any value of $\gamma$ in the range $0\leq \gamma \leq 2$, except $\gamma\neq(\zeta_1+4\alpha \zeta_1)$ and $\alpha\neq-1/2 $ as
\begin{equation}
H=\frac{e^{\frac{3
(1+2\alpha)(1+4\alpha)\zeta_{0}}{2(1+4\alpha-\alpha\gamma)}t}}{c_{6}+\frac{(\gamma-\zeta_{1}-4\alpha\zeta_{1})}
{(1+4\alpha)\zeta_{0}}e^{\frac{3
 (1+2\alpha)(1+4\alpha)\zeta_{0}}{2(1+4\alpha-\alpha\gamma)}t}},
\end{equation}
 where $c_{6}$ is constant of integration. The scale factor in terms of t is given by\\
\begin{equation}
a=c_{7}\left[c_{6}+\frac{(\gamma-\zeta_{1}-4\alpha\zeta_{1})}
{(1+4\alpha)\zeta_{0}}\;e^{\frac{3}{2}\frac{(1+2\alpha)(1+4\alpha)\zeta_{0}}{(1+4\alpha-\alpha\gamma)}\;t}
\right]^{\frac{2(1+4\alpha-\alpha\gamma)}{3(\gamma-\zeta_{1}-4\alpha\zeta_{1})(1+2\alpha)}},
\end{equation}
where $c_{7 }>0$ is a constant of integration. This scale factor may be rewritten as \\
\begin{equation}
a=a_{0}\left[1+\frac{H_{0}(\gamma-\zeta_{1}-4\alpha\zeta_{1})}
{(1+4\alpha)\zeta_{0}}\;\left(e^{\frac{3}{2}\frac{(1+2\alpha)(1+4\alpha)\zeta_{0}}{(1+4\alpha-\alpha\gamma)}\;t}-1\right)
\right]^{\frac{2(1+4\alpha-\alpha\gamma)}{3(\gamma-\zeta_{1}-4\alpha\zeta_{1})(1+2\alpha)}}.\\
\end{equation}
\indent The energy density $\rho$ and bulk viscosity $\zeta$ can be calculated as
\begin{eqnarray}
\rho &=&\frac{3H_{0}}{(1+4\alpha-\alpha \gamma)}\frac{e^{\frac{3}{2}\frac{(1+2\alpha)(1+4\alpha)\zeta_{0}}{(1+4\alpha-\alpha\gamma)}\;t}}
{\left[1+\frac{(\gamma-\zeta_{1}-4\alpha\zeta_{1}) H_{0}}
{(1+4\alpha)\zeta_{0}}\;\left(e^{\frac{3}{2}\frac{(1+2\alpha)(1+4\alpha)\zeta_{0}}{(1+4\alpha-\alpha\gamma)}\;t}-1\right)\right]}\nonumber\\
& &\left[\frac{(1-\alpha\zeta_{0})H_{0}\;e^{\frac{3}{2}\frac{(1+2\alpha)(1+4\alpha)\zeta_{0}}
{(1+4\alpha-\alpha\gamma)}\;t}}{1+\frac{(\gamma-\zeta_{1}-4\alpha\zeta_{1}) H_{0}}
{(1+4\alpha)\zeta_{0}}\;\left(e^{\frac{3}{2}\frac{(1+2\alpha)(1+4\alpha)\zeta_{0}}{(1+4\alpha-\alpha\gamma)}\;t}-1\right)}-\alpha\; \zeta_{0}\right],
\end{eqnarray}
\begin{equation}
\zeta= \zeta_{0}+\zeta_{1}\left[ \frac{H_{0}\;e^{\frac{3
(1+2\alpha)(1+4\alpha)\zeta_{0}}{2(1+4\alpha-\alpha\gamma)}t}}{1+\frac{H_{0}(\gamma-\zeta_{1}-4\alpha\zeta_{1})}
{(1+4\alpha)\zeta_{0}}\left(e^{\frac{3
(1+2\alpha)(1+4\alpha)\zeta_{0}}{2(1+4\alpha-\alpha\gamma)}t}-1\right)}\right].
\end{equation}
With the constraint given in (47) we get a big rip singularity at
\begin{equation}
t_{br}=\frac{2(1+4\alpha-\alpha \gamma )}{3(1+2\alpha)(1+4\alpha)\zeta_{0}}\;ln \left(1+\frac{(1+4\alpha)\zeta_{0}}{ H_{0}(\zeta_1+4\alpha\zeta_1-\gamma)}\right).
\end{equation}
In this case the deceleration parameter is given by\\
\begin{equation}
q=\frac{\frac{3}{2}\frac{(1+2\alpha)}{(1+4\alpha-\alpha \gamma)}[(\gamma-\zeta_{1}-4\alpha\zeta_{1})-\frac{(1+4\alpha)\zeta_{0}}{H_{0}}]}
{e^{\frac{3}{2}\frac{(1+2\alpha)(1+4\alpha)\zeta_{0}}{(1+4\alpha-\alpha\gamma)}\;t}}-1,
\end{equation}\\
\noindent which shows that $q$ is time-dependent. Therefore, we study the variation of $q$ with bulk viscous coefficient $\zeta_{0}+\zeta_{1}H$ for various ranges of $\alpha$ in different phases of evolution of the universe, which are summarized in tables 5-8.

\vspace{0.3cm}

\begin{center}
\textbf{Table. 5} : \textsf{Variation of $q$ for $\gamma=2/3$.}\\
\vspace{0.2cm}
\footnotesize{\begin{tabular}{|c|c|c|c|}
  \hline
  Range of $\alpha $& Constraints on $\zeta_{0}$ and $\zeta_{1}$ & $q$ &Evolution of Universe \\ \hline
  $\alpha > 0$ & for all $\zeta_{0}>0 \;{\text{and}}\; \zeta_{1}>0$ & negative & accelerated expansion  \\\hline
  $-0.25<\alpha<0$  & $0<(\zeta_{0}+H_{0}\zeta_{1})<\frac{-8\alpha H_{0}}{9(1+2\alpha)(1+4\alpha)}$ & +ve to -ve & transition from dec. to acc. \\
  & $(\zeta_{0}+H_{0}\zeta_{1})\geq\frac{-8\alpha H_{0}}{9(1+2\alpha)(1+4\alpha)}$ & negative &  accelerated expansion  \\ \hline
  $-0.30<\alpha\leq-0.25$ & for all $\zeta_{0}>0$ and $\zeta_{1}>0$ & positive & decelerated expansion  \\ \hline
$-0.50\leq \alpha<-0.30$ & for all $\zeta_{0}>0$ and $\zeta_{1}>0$ & negative & accelerated  expansion  \\ \hline
 $\alpha<-0.50$  & $0<(\zeta_{0}+H_{0}\zeta_{1})<\frac{-8\alpha H_{0}}{9(1+2\alpha)(1+4\alpha)}$ & -ve to +ve & transition from acc. to dec.\\
  &$(\zeta_{0}+H_{0}\zeta_{1})\geq\frac{-8\alpha H_{0}}{9(1+2\alpha)(1+4\alpha)}$ & positive
& decelerated expansion\\ \hline
\end{tabular}}
\end{center}

\vspace{0.3cm}

\begin{center}
\textbf{Table. 6} : \textsf{Variation of $q$ for $\gamma=4/3$.}\\
\vspace{0.2cm}
\footnotesize{\begin{tabular}{|c|c|c|c|}
  \hline
  Range of $\alpha $ & Constraints on $\zeta_{0}$ and $\zeta_{1}$ & $q$ &Evolution of Universe \\ \hline
  $\alpha > -0.25$ ($\alpha\neq 0$) & $0<(\zeta_{0}+H_{0}\zeta_{1}) <\frac{2(3+4\alpha)H_{0}} {9(1+2\alpha)(1+4\alpha)} $ & +ve to -ve & transition from dec. to acc. \\
  & $(\zeta_{0}+H_{0}\zeta_{1})\geq \frac{2(3+4\alpha)H_{0}}{9(1+2\alpha)(1+4\alpha)} $ & negative & accelerated expantion\\ \hline
 $-0.375<\alpha\leq-0.25$ & for all $0<(\zeta_{0}+H_{0}\zeta_{1})$ & positive & decelerated expansion  \\ \hline
$-0.50\leq \alpha<-0.375$ & for all $0<(\zeta_{0}+H_{0}\zeta_{1})$ & negative & accelerated  expansion  \\ \hline
$\alpha<-0.50$  & $0<(\zeta_{0}+H_{0}\zeta_{1}) <\frac{2(3+4\alpha)H_{0}} {9(1+2\alpha)(1+4\alpha)} $ & -ve to +ve & transition from acc. to dec.\\
  &$(\zeta_{0}+H_{0}\zeta_{1})\geq\frac{2(3+4\alpha)H_{0}}{9(1+2\alpha)(1+4\alpha)}$ & positive
& decelerated expansion\\ \hline
\end{tabular}}
\end{center}

\vspace{0.3cm}

\begin{center}
\textbf{Table. 7} : \textsf{Variation of $q$ for $\gamma=1$.}\\
\vspace{0.2cm}
\footnotesize{\begin{tabular}{|c|c|c|c|}
  \hline
  Range of $\alpha $ & Constraints on $\zeta_{0}$ and $\zeta_{1}$ & $q$ &Evolution of Universe \\ \hline
  $\alpha > -0.25$ ($\alpha\neq 0$) & $0<(\zeta_{0}+H_{0}\zeta_{1})<\frac{H_{0}}{3(1+2\alpha)(1+4\alpha)} $ & +ve to -ve & transition from dec. to acc. \\
  & $(\zeta_{0}+H_{0}\zeta_{1})\geq \frac{H_{0}}{3(1+2\alpha)(1+4\alpha)} $ & negative & accelerated expantion\\ \hline
 $-0.33<\alpha\leq-0.25$ & for all $\zeta_{0}>0$ and $\zeta_{1}>0$ & positive & decelerated expansion  \\ \hline
$-0.50\leq \alpha<-0.33$ & for all $\zeta_{0}>0$ and $\zeta_{1}>0$ & negative & accelerated  expansion  \\ \hline
$\alpha<-0.50$  & $0 < (\zeta_{0}+H_{0}\zeta_{1})<\frac{H_{0}}{3(1+2\alpha)(1+4\alpha)} $ & -ve to +ve & transition from acc. to dec.\\
  &$(\zeta_{0}+H_{0}\zeta_{1})\geq\frac{H_{0}}{3(1+2\alpha)(1+4\alpha)}$ & positive
& decelerated expansion\\ \hline
\end{tabular}}
\end{center}

\vspace{0.3cm}

\begin{center}
\textbf{Table. 8} : \textsf{Variation of $q$ for $\gamma=0$.}\\
\vspace{0.2cm}
\footnotesize{\begin{tabular}{|c|c|c|c|}
  \hline
  Range of $\alpha $ & Constraints on $\zeta_{0}$ and $\zeta_{1}$ & $q$ &Evolution of Universe \\ \hline
  $\alpha \geq -0.50$ ($\alpha\neq 0$)  & for all $\zeta_{0}>0\;{\text{and}}\; \zeta_{1}>0$  & negative & accelerated  expansion  \\ \hline
$\alpha<-0.50$  & $0<(\zeta_{0}+H_{0}\zeta_{1})<-\frac{2H_{0}}{3(1+2\alpha)} $ & -ve to +ve & transition from acc. to dec.\\
  &$(\zeta_{0}+H_{0}\zeta_{1})\geq-\frac{2H_{0}}{3(1+2\alpha)}$ & positive & decelerated expansion\\ \hline
\end{tabular}}
\end{center}

\noindent We observe from the above tables 5-8 that the universe accelerates through out the evolution when $\alpha > 0$ for any positive values of $\zeta_{0}$ and $ \zeta_{1} $ in inflationary phase where as it shows transition from decelerated phase to accelerated phase when $\alpha > -0.25$ for smaller values of $(\zeta_{0}+H_{0}\zeta_{1})$ and shows acceleration for larger values of $(\zeta_{0}+H_{0}\zeta_{1})$ in radiation and matter-dominated phases. It is due to the fact that the larger values of $\zeta_{0}$ or $\zeta_{1}$ or both  make the effective pressure more negative to accelerate the universe through out the evolution. We find that the universe decelerates in $-0.30<\alpha\leq-0.25$ for $\gamma=2/3$, in $-0.375<\alpha\leq-0.25$ for $\gamma=4/3$ and in $-0.33<\alpha\leq-0.25$ for $\gamma=1$ for all $\zeta_{0}>0$ and $\zeta_{1}>0$. Further, we find that the universe accelerates in $-0.50\leq \alpha<-0.30$ for $\gamma=2/3$, in $-0.50\leq \alpha<-0.375$ for $\gamma=4/3$ and in $-0.50\leq \alpha<-0.33$ for $\gamma=1$ for all $\zeta_{0}>0$ and $\zeta_{1}>0$. For $\gamma=0$ and $\alpha\geq -0.50$, we observe that universe shows accelerated expansion for positive values of $\zeta_{0}$ and $ \zeta_{1}$.
  When $\alpha <-0.50$, the model shows transition from acceleration to deceleration for small values of $\zeta_{0}+ H_{0} \zeta_{1}$  and decelerates for large values of $\zeta_{0}+ H_{0} \zeta_{1}$ for all phases.\\
\indent For $\zeta_{1}(1+4\alpha)> \gamma$, where the model has a big rip singularity, $q$ is always negative for $\alpha>0$ during any cosmic time.  We also see from (40) and (49) that $\dot H=0$ for $\zeta_{1}(1+4\alpha)= \gamma$ or $\alpha=-1/2$, which shows de-Sitter expansion of scale factor in both cases. The solutions for $\gamma=0$ can be obtained directly from (43) and (52), and  the big rip singularity can be observed at $t_{br}=2/[3(1+2\alpha)\zeta_1H_{0}]>t_0$ in case I and $t_{br}=\frac{2}{3(1+2\alpha)\zeta_{0}}\;ln \left(1+\frac{\zeta_{0}}{ H_{0}\zeta_1}\right)>t_0$ in case II, respectively.\\

\noindent \textbf{5 Conclusion}\\

\noindent Harko et al.[20] proposed the modified f(R,T) theory of gravity with perfect fluid (dust model) and obtained a power-law singular solution by assuming $f(R,T)=R+2\alpha T$, where $\alpha$ is constant. Later on, many authors [21, 22-26, 55] studied this modified theory with a combination of perfect fluid and scalar field by reconstructing $f(R,T)$ gravity with the same form of $f(R,T)$. Most treatises on modified gravity, as well as on standard gravity, assume the cosmic fluid to be ideal, i.e., non-viscous. But many works as discussed in section 1 describe the evolution of the universe with dissipative processes due to viscosity.  Cosmic bulk viscosity is a viable candidate to explain early and late time expansion of the universe. Therefore, in this paper we have explored the evolution of the universe driven by a kind of viscous fluid by assuming general form of bulk viscous term. We have extensively studied the effects of viscous fluid in $f(R,T)$ gravity within the framework of a flat FRW model. We have investigated the dissipative processes  of the standard Eckart theory of relativistic irreversible thermodynamics. We have discussed the expansion history of the universe with and without bulk viscosity. Power-law, exponential and superinflation non- singular solutions have been obtained according to the choices of $\zeta$. We have also found cosmological solutions which exhibit big rip singularity under certain constraints. Therefore, the negative pressure generated by the bulk viscosity can not avoid the dark energy of the universe to be phantom. It is to be noted that we have discussed the various phases and their possible transitions for all possible range of $\alpha$ with $\zeta_0$ and $\zeta_1$ , which have not been studied in the past cited works. It contains many new solutions like power-law, exponential and superinflationary scale factors by assuming same form of $f(R,T)$. We have obtained both constant and time-dependent deceleration parameters which describe the decelerated / accelerated and transition from decelerated to accelerated  phase. As we have considered all possible positive and negative ranges of $\alpha$ with the inclusion of viscous term in $f(R,T)$ theory which is of physical intersts, one may see that our results are relatively more generalized from the past works. We summarize the results of each section one by one as follows:\\
\indent In case of perfect fluid distribution as presented in section 4.1, a power-law expansion for the scale factor has been obtained for $\gamma\neq 0$. It has found that the energy density decreases with time and the deceleration parameter is constant throughout the evolution. For $\gamma=0$,i.e., when $p=-\rho$, we have de-Sitter solution of scale factor where both $p$ and $\rho$ are constants. For $\gamma <0$, there is a big rip singularity at a finite value of cosmic time describing the phantom cosmology.\\
{\begin{center}
\begin{tabular}{cc}
\begin{minipage}{200pt}
\frame{\includegraphics[width=200pt]{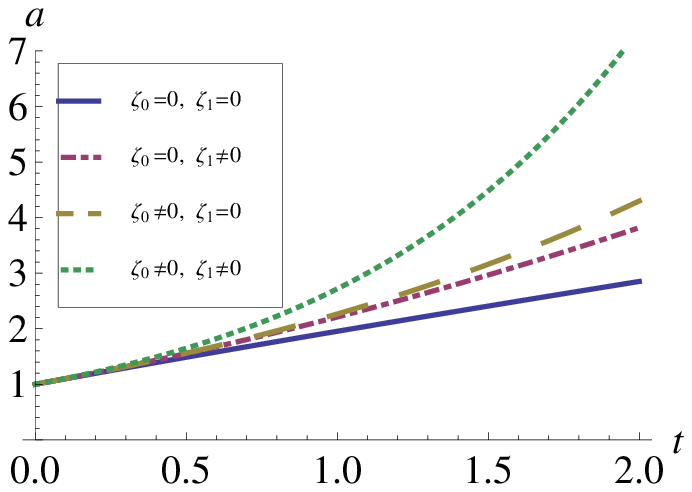}}
\center{\footnotesize \textbf{Fig. 1(a)} Scale factor in terms of cosmic time for $\alpha>0$. Here $\gamma=\alpha= a_{0}=H_{0}=1$ and $\zeta_{0}=\zeta_{1}=0.1$ . }
\end{minipage}&\begin{minipage}{200pt}
\frame{\includegraphics[width=200pt]{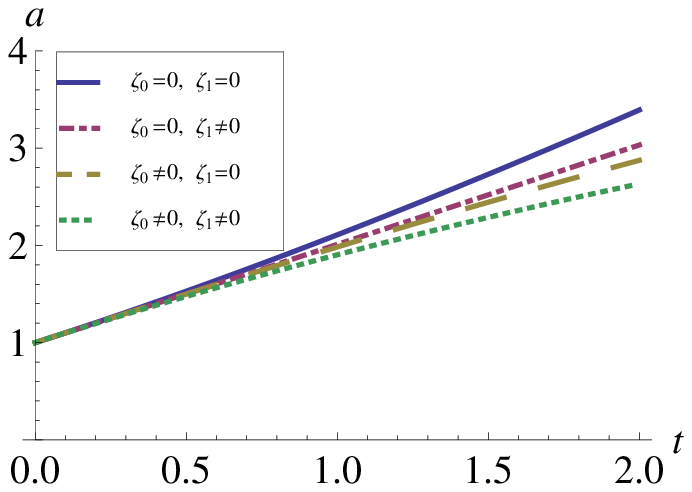}}
\center{\footnotesize \textbf{Fig. 1(b)} Scale factor in terms of cosmic time for $\alpha<0$. Here $\alpha=-1$, $\gamma= a_{0}=H_{0}=1$ and $\zeta_{0}=\zeta_{1}=0.1$ }
\end{minipage}
\end{tabular}
\end{center}}
\indent In case of viscous cosmology, we have explored bulk viscous model composed by perfect fluid with a general bulk viscous form $\zeta=\zeta_{0}+\zeta_{1}H$. We have discussed three different cases depending on the composition of the bulk viscosity. In case of constant coefficient of bulk viscosity, the scale factor varies exponentially for $\gamma \neq 0$, which avoids the big-bang singularity. We have observed that the constant viscous term generates the time-dependent deceleration parameter, which describes the different phases of the universe and transition to accelerated phase. We have shown the variation of $q$ and the corresponding evolution of the universe in tables 1-4 for various ranges of $\alpha$. In this case, we also have the big rip singularity at finite value of cosmic time for $\gamma<0$. When $\gamma=0$, a superinflation for the scale factor has been found whereas it has de Sitter solution in non-viscous case.\\
\indent In case where the bulk viscosity is proportional to Hubble parameter, we have obtained a power-law expansion of the scale factor similar to the perfect fluid model. The energy density varies inversely as the square of the cosmic time whereas the bulk viscosity decreases linearly with time. In this case the value of $q$ is constant and the sign of $q$ depends on $\gamma$, $\alpha$ and $\zeta_1$. This form of bulk viscosity generates a big rip singularity when the constraint given in (47) holds. \\
\indent In case of $\zeta=\zeta_0+\zeta_1 H$, we again have obtained exponential form of expansion for the scale factor which is similar to the form with constant bulk viscous term. It has been observed that the deceleration parameter is time-dependent which can explain the expansion history of the universe and the transition to accelerated or decelerated phase (see tables 5-8).\\
\indent The behavior of scale factor with cosmic time has been shown in Figures 1a and 1b for different models with the particular choices of bulk viscosity. It has been observed that the scale factor increases rapidly in all viscous models as compared to the non-viscous model(perfect fluid model) for any  positive values of $\alpha$ as shown in figure 1a. The rate of expansion depends on the bulk viscous coefficient. The scale factor varies very close to non-viscous model when bulk viscous coefficient is assumed to be small. Figure 1a shows that the expansion of scale factors deviate more rapidly from the perfect fluid expansion rate for larger bulk viscous coefficient and the rate of expansion is very fast when the bulk viscosity is considered in the form $\zeta=\zeta_0+\zeta_1H$. The behavior of scale factor for different models is reverse for $\alpha<0$ as shown in figure 1b. The variation of deceleration parameter with cosmic time has been shown in figure 2 for different models. The models with $\zeta=0$ (perfect fluid) and $\zeta=\zeta_1 H$, where the scale factor vary as power-law of cosmic time, the values of $q$ are constant. On the other hand, in models $\zeta=\zeta_0$ and $\zeta=\zeta_0+\zeta_1H$, where the scale factor varies exponentially, $q$ is time-dependent and transitions from positive to negative values are observed.\\
\begin{figure}
\centering
  \includegraphics[width=6.5cm, height=4.5cm]{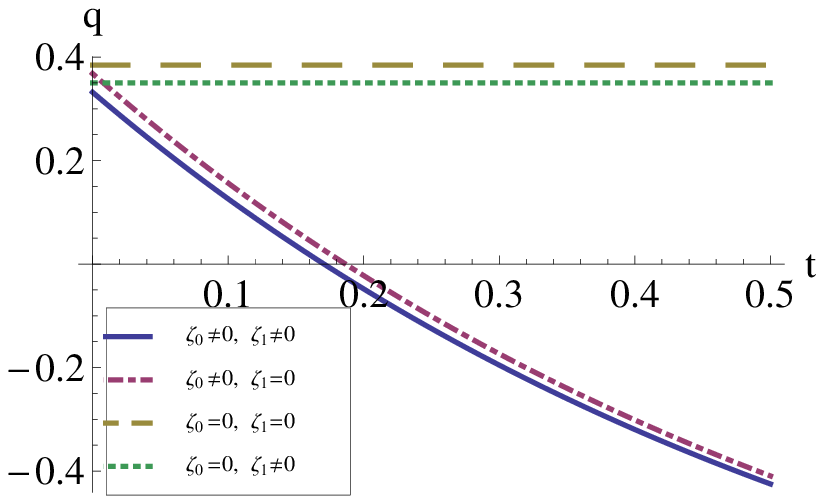}\\
 {\footnotesize \textbf{Fig. 2} Deceleration parameter in terms of cosmic time for $\alpha>0$. Here $\gamma=H_{0}=1, \alpha=0.1, \zeta_{0}=0.01$ and $\zeta_{1}=0.02$.}
\end{figure}
\indent In conclusion we emphasize that perfect fluid is just a limiting case of a general viscous media that is more practical in the astrophysical sense.  Therefore, it is worthy to study the early and late time evolution of the universe with viscous fluid in $f(R,T)$ theory of gravity which describes the evolution of the universe with different range of $\alpha$. \\

\noindent \textbf{Acknowledgement}\\

\noindent The authors thank the referee for his useful suggestions and comments to improve the manuscript. One of author (PK) is grateful to University Grants Commission, India for financial support under Junior Research Fellowship scheme.\\

\noindent \textbf{References}\\

\noindent 1. H.A. Buchdahl,  Mon. Not. R. Astro. Soc. {\bf150}, 1 (1970)\\
\noindent 2. A.A. Starobinsky, Phys. Lett. B {\bf91}, 99 (1980)\\
\noindent 3. S. Nojiri, S.D. Odintsov, Phys. Rev. D {\bf68}, 123512 (2003)\\
\noindent 4. T.P. Sotiriou, V. Faraoni, Rev. Mod. Phys. {\bf82}, 451 (2010)\\
\noindent 5. S. Capozziello, M. De Laurents, Phys. Rept. {\bf 509}, 167 (2011)\\
\noindent 6. T. Clifton, P.G. Ferreira, A. Padilla, C. Skordis, Phys. Rept. {\bf513}, 1 (2012)\\
\noindent 7. S. Nojiri, S.D. Odintsov, Phys. Lett. B {\bf562}, 147 (2003)\\
\noindent 8. S. Nojiri, S.D. Odintsov, Phys. Lett. A {\bf19}, 627 (2004)\\
\noindent 9. S. Nojiri, S.D. Odintsov, Int. J. Geom. Meth. Mod. Phys. {\bf4}, 115 (2007)\\
\noindent 10. S.M. Carroll,  V. Duwuri, M. Trodden, M.S. Turner, Phys. Rev. D {\bf 70}, 043528 (2004)\\
\noindent 11. T. Chiba, A.L. Erichcek, Phys. Rev. D {\bf75}, 124014 (2007)\\
\noindent 12. A.A. Starobinsky, J. Exp. Theor. Phys. Lett. {\bf86}, 157 (2007)\\
\noindent 13. S. Nojiri, S.D. Odintsov, Phys. Rep. {\bf505}, 59 (2011); K. Bamba, S. Capozziello, S. Nojiri, \indent S.D. Odintsov, Astrophys. Space Sci. {\bf342}, 155 (2012), [arXiv: gr-qc/1205.3421]\\
\noindent 14. K. Bamba, S. Nojiri, S.D. Odintsov, J. Cosmol. Astropart. Phys. {\bf0180}, 045 (2008)\\
\noindent 15. S. Nojiri, S.D. Odintsov, M. Sami, Phys. Rev. D {\bf74}, 046004 (2006)\\
\noindent 16. R. Femaro, F. Fiorini, Phys. Rev. D {\bf75}, 084031 (2007)\\
\noindent 17. O. Bertolami, C.G. Bochmer, T. Harko, F.S.N. Lobo, Phys. Rev. D {\bf75}, 104016 (2007)\\
\noindent 18. T. Harko,  Phys. Lett. B {\bf 669}, 376 (2008)\\
\noindent 19. T. Harko, F.S.N. Lobo, Eur. Phys. J. C {\bf70}, 373 (2010)\\
\noindent 20. T. Harko, F.S.N. Lobo, S. Nojiri, S.D. Odintsov,  Phys. Rev. D {\bf 84}, 024020 (2011),\\ \indent [arXiv:gr-qc/1104.2669].\\
\noindent 21. M.J.S. Houndjo, Int. J. Mod. Phys. D 21, 1250003 (2012), [arXiv: astro-ph/1107.3887]; \indent M.J.S. Houndjo, O.F. Piattella, Int. J. Mod. Phys. D 21, 1250024 (2012), [arXiv: gr-\indent qc/1111.4275]; M.J.S. Houndjo, C.E.M. Batista, J.P. Campos, O.F. Piattella,  Canadian \indent J. Phys. {\bf91}, 548 (2013), [arXiv:gr-qc/1203.6084]\\
\noindent22. F.G. Alvarenga, M.J.S. Houndjo, A.V. Monwanou, J.B. Chobi Oron,  J. Mod. Phys. \\
\indent {\bf4}, 130 (2013), [arXiv:gr-qc/1205.4678]\\
\noindent 23. A. Pasqua, S. Chattopadhyay, I. Khomenkoc,  Canadian J. Phys. {\bf 91}, 632  (2013)\\
\noindent 24. M. Sharif, M. Zubair,  J. Cosmol. Astropart. Phys. {\bf 21}, 28 (2012), [arXiv:gr-qc/1204.0848]\\
\noindent 25. T. Azizi,  Int J. Theor. Phys. {\bf 52}, 3486 (2013), [arXiv:gr-qc/1205.6957]\\
\noindent 26. S. Chakraborty,  Gen. Relativ. Grav. {\bf45}, 2039 (2013), [arXiv:gen-ph/1212.3050]\\
\noindent 27. C.P. Singh and V. Singh,  Gen. Relativ. Grav.} {\bf46}, 1696 (2014)\\
\noindent 28. A.G. Riess, et al.,  Astron. J.  {\bf116}, 1009 (1998)\\
\noindent 29. S. Perlmutter, et al.,  Astrophys. J. {\bf517}, 565 (1999)\\
\noindent 30. A.G. Riess, et al.,  Astrophys. J. {\bf607}, 665 (2004)\\
\noindent 31. A.G. Riess, et al.,  Astrophys. J. {\bf659}, 98 (2007)\\
\noindent 32. E. Komatsu, et al.,  Astrophys. J. Suppl. {\bf192}, 18 (2011)\\
\noindent 33. C.L. Bennett, et al.,  Astrophys. J. {\bf148}, 1 (2003)\\
\noindent 34. K. Abazajian, et al., Astron. j. {\bf 128}, 502 (2004)\\
\noindent 35. M. Tegmark, et al.,  Phys. Rev. D {\bf 69}, 103501 (2004)\\
\noindent 36. W. Misner,  Astrophysical J. {\bf151}, 431 (1968); W. Israel and J.N. Vardalas, Nuovo \indent Cimento Lett. {\bf4}, 887 (1970); Z. Klimek, Postepy,  Astron. {\bf19}, 165 (1971); G.L. Murphy, \indent Phys. Rev. D  {\bf8}, 4231 (1973);  V.A. Belinskii and I. M. Kalatnikov, Pisma Zh. Eksp.  \indent Tekhn. Fiz. {\bf 21}, 223 (1974)\\
\noindent 37. L. Diosi, B. Keszthelyi, B. Lukacs, and G. Paal, Acta Phys. Pol. B {\bf15}, 909 (1984); \indent I.Waga, R.C. Falcao, and R. Chanda, Phys. Rev. D {\bf33}, 1839 (1986); J.D. Barrow, \indent Phys. Lett. {\bf180}, 335 (1986); J. D. Barrow, Nucl. Phys. B {\bf380}, 743 (1988); \\
\noindent 38. W. Zimdahl, D.J. Schwarz, A.B. Balakin, and D. Pavon, Phys. Rev. D {\bf64}, 063501 (2001)\\
\noindent 39. H. Velten and D. J. Schwarz, Phys. Rev. D {\bf86}, 083501 (2012)\\
\noindent 40. M.R. Setare and A. Sheykhi, Int. J. Mod. Phys.D {\bf19}, 1205 (2010); Mauricio Catal- \indent do, Norman Cruz and Samuel Lepe, Phys. Lett. B {\bf619}, 5 (2005); I. Brevik and O. \indent Gorbunova, Gen. Relativ. Grav. {\bf37}, 2039 (2005); Jean-Sebastien Gagnon and Julien \indent Lesgourgues, J. Cosmol. Astropart. Phys. {\bf09},  026 (2011); I Brevik, E. Elizalde and \indent S.D. Odintsov, Phys. Rev.D {\bf84}, 103508 (2011),[arxiv:hep-th/1107.4642]\\
\noindent 41. B. Li and J.D. Barrow, Phys. Rev. D {\bf79}, 103521 (2009); W.S. Hip´olito-Ricaldi, H.E.S. \indent Velten and W. Zimdahl, J. Cosmol. Astropart. Phys.{\bf06}, 016 (2009); W.S. Hip´olito-\indent Ricaldi, H.E.S. Velten  and  W. Zimdahl, Phys.Rev. D {\bf82}, 063507 (2010); A. Montiel and \indent N. Bretn, J. Cosmol. Astropart. Phys. {\bf08}, 023 (2011); J.C. Fabris, P.L.C. de Oliveira \indent  and H.E.S. Velten, Eur.Phys.J. C {\bf71}, 1773 (2011); H. Velten and D.J. Schwarz, J. Cos-\indent mol. Astropart. Phys.  {\bf09}, 016 (2011).\\
\noindent 42. Ming-Guang Hu,  Xin-He Meng,  Phys. Lett. B  {\bf635}, 186 (2006), [arXiv:astro-ph/0511615];  \indent Jie Ren and Xin-He Meng,  Phys. Lett. B {\bf 633},  1 (2006); Jie Ren and Xin-He Meng, Phys. \indent Lett. B  {\bf636}, 5 (2006)\\
\noindent 43. J.S. Gagnon, J. Lesgourgues, J. Cosmol. Astropart. Phys. {\bf 09}, 026 (2011), \indent [arXiv:astro-\indent ph/1107.1503v2]\\
\noindent 44. A.A. Sen, et al.,  Phys. Rev. D {\bf 63}, 107501 (2001)\\
\noindent 45. M.K. Mak and T. Harko, {\it Int. J. Mod. Phys. D} {\bf 12}, 925 (2003); J.C. Fabris, S.V.B. \indent Goncalves, R. de S´a Ribeiro,   Gen. Relativ. Gravit. {\bf38}, 495 (2006)\\
\noindent 46. C.P. Singh, S. Kumar, A. Pradhan,  Class. Quantum Grav. {\bf24}, 455 (2007)\\
\noindent 47. C.P. Singh, Pramana J. Phys. {\bf 72}, 429 (2008)\\
\noindent 48. X.H. Meng, X. Dou, Commun. Theor. Phys. {\bf52}, 377 (2009), [arXiv:astro-ph/0812.4904]\\
\noindent 49. S. Weinberg,  Astrophys. J. {\bf168}, 175 (1972)\\
\noindent 50. I. Brevik,  Phys. Rev. D {\bf65}, 127302 (2002)\\
\noindent 51. C. Eckart,  Phys. Rev. {\bf58}, 919 (1940)\\
\noindent 52. Xin-He Meng, Jie Ren, Ming-Gaung Hu,  Commun. Theor. Phys. {\bf47}, 379 (2007)\\
\noindent 53. A. Avelino, U. Nucamendi, J. Cosmol. Astropart. Phys. {\bf9}, 1008 (2010),\\ \indent[arXiv:gr-qc/1002.3605]\\
\noindent 54. {\O}. Gr{\o}n,  Astrophys. space Sci. {\bf173}, 191 (1990)\\
\noindent 55. M. Jamil, D. momeni, M. raza and R. Myrzakulov, Eur. Phys. J. C {\bf72}, 1999 (2012); M. \indent Jamil, D. Momeni and R. Myrzakulov, Chin. Phys. Lett. {\bf29}, 109801 (2012)\\

\end{document}